\documentclass[%
 reprint,
 amsmath,amssymb,
 superscriptaddress,
 aps
]{revtex4-2}

\usepackage{graphicx}
\usepackage{dcolumn}
\usepackage{bm}
\usepackage{hyperref}
\hypersetup{
    colorlinks = true,
    citecolor=blue,
	urlcolor=blue,
	linkcolor=blue
}
\usepackage[inline]{enumitem}
\begin{document}

\title{Unified definition of exciton coherence length for exciton-phonon coupled molecular aggregates}%
\author{Tong Jiang}
\affiliation{MOE Key Laboratory of Organic OptoElectronics and Molecular
Engineering,
Department of Chemistry, Tsinghua University, Beijing
100084,
People's Republic of China\\
}
\author{Jiajun Ren}
\affiliation{MOE Key Laboratory of Theoretical and Computational Photochemistry,
College of Chemistry, Beijing Normal University, Beijing 100875, People’s Republic of China
}
\author{Zhigang Shuai}%
\email{zgshuai@tsinghua.edu.cn}
\affiliation{MOE Key Laboratory of Organic OptoElectronics and Molecular
Engineering,
Department of Chemistry, Tsinghua University, Beijing
100084,
People's Republic of China\\
}
\affiliation{School of Science and Engineering, The Chinese University of Hong Kong, Shenzhen, Guangdong 518172, People's Republic of China}
\date{\today}

\begin{abstract}
Exciton coherence length (ECL) is an essential concept to characterize the nature of exciton in molecular aggregates for photosynthesis, organic photovoltaics, and light-emitting diodes. 
ECL has been defined in a number of ways through the variance or purity of the electronic reduced density matrix. 
However, we find that these definitions fail to present a monotonic relationship with respect to the exciton radiative decay efficiency as it should be when exciton-phonon couplings are taken into accounts. We propose a unified definition of ECL by virtue of sum rule of oscillator strengths. Using the numerically accurate time-dependent matrix product states formalism applied to Frenkel-Holstein models for molecular aggregates in both one- and two-dimensional system, we find our ECL definition exhibits a monotonic relationship with respect to the radiative efficiency and can serve as an efficient and unified description for exciton coherence. We further predict that the two-dimensional aggregates can display maximum superradiance enhancement (SRE) at finite temperature, different from the previous knowledge of SRE-$1/T$ behavior.
\end{abstract}

\maketitle
Molecular aggregates have received great attentions due to their promising applications in biological photosynthesis~\cite{cao2020quantum,jang2018delocalized}, organic light-emitting, lasers, energy conversion such as photovoltaics and thermoelectrics, and field effect transistors~\cite{wei2020overcoming,jin2018long,Liu2021, ostroverkhova2016organic}, \textit{etc.}
Great progresses in experiments and theories have deepened the understanding of the excited states of molecular aggregates --- excitons~\cite{dimitriev2022dynamics,scholes2017using}.
Exciton in molecular aggregates is the collective excitation consisting of a linear combination of local excitations, where molecules subject to photo absorption exchange excitation energy through transition dipole-dipole interaction (excitonic coupling)~\cite{zhang2016visualizing}.
When the local dipole oscillation is instructive, the dipoles add up coherently and thus achieve superradiance in J aggregate~\cite{gross1982superradiance,spano1990temperature, heijs2005decoherence,eisfeld2017superradiance,hestand2018expanded}, while destructive phase between dipoles leads to fluorescence quenching in H aggregate~\cite{kasha1963energy,kasha1965exciton,hestand2018expanded}.
Ideal exciton is fully coherent over the entire domain for aggregates composed of $N$ \textit{disorder-free rigid} molecules at zero temperature, leading to $N$ times enhancement of the fluorescence rate for the former case and complete quench for the latter.
In reality, due to the soft nature, organic $\pi$ conjugated molecules undergo structure reorganizations upon excitation and the exciton is spatially localized to a subset of the aggregate in time scales of sub-picoseconds or picoseconds due to the coupling to high frequency stretching modes~\cite{hestand2018expanded,park2018excited} or low frequency torsional modes~\cite{binder2018conformational,deutsch2020geometry}.
The exciton coherence length (ECL) measures the spatial distance over which the exciton is delocalized, which is of considerable interest for its determining impact in optical property, especially the radiative efficiency referring to the ratio of aggregate's radiative rate over that of isolated monomer.

Unfortunately, ECL has been long poorly defined for electron-phonon coupled molecular aggregates.
One way to obtain the exciton coherence is calculating the emission intensity ratio between 0-0 and 0-1 peak~\cite{spano2011vibronic}, which is restricted to J aggregate consisting of molecules coupled to one high-frequency intramolecular mode.
The most relevant physical quantity to exciton coherence is the exciton reduced density matrix, obtained by tracing the phonon degrees of freedom of the total density matrix.
The reduced density matrix depicts the system at thermal equilibrium and serves as the initial state for quantum dynamics simulations of open quantum systems.
Multiple definitions of ECL have been proposed to understand the coherence and its relation to radiative efficiency~\cite{meier1997polarons,kuhn1997pump,sarovar2010quantum,scholes2020limits}.
However, their generalization to complex cases is questioned that different definitions of ECL lead to even qualitative differences for the temperature dependence of coherence~\cite{moix2012equilibrium}. 
Therefore the use of these defintions should be cautioned.
It is the purpose of this Letter to give a unified definition of ECL that connects rigorously with the radiative efficiency.

Despite the importance of exciton reduced density matrix, very few numerically exact methods have been employed to calculate it.
We recently developed an algorithm that combined the imaginary time evolution~\cite{zwolak2004mixed,verstraete2004matrix} with the matrix product states (MPS) to calculate the thermal equilibrium density matrix of the high dimensional exciton-phonon systems~\cite{ren2018time}, which is applied to solve the ECL definition problem here.
\begin{figure}
\centering
\includegraphics[width=0.35\textwidth]{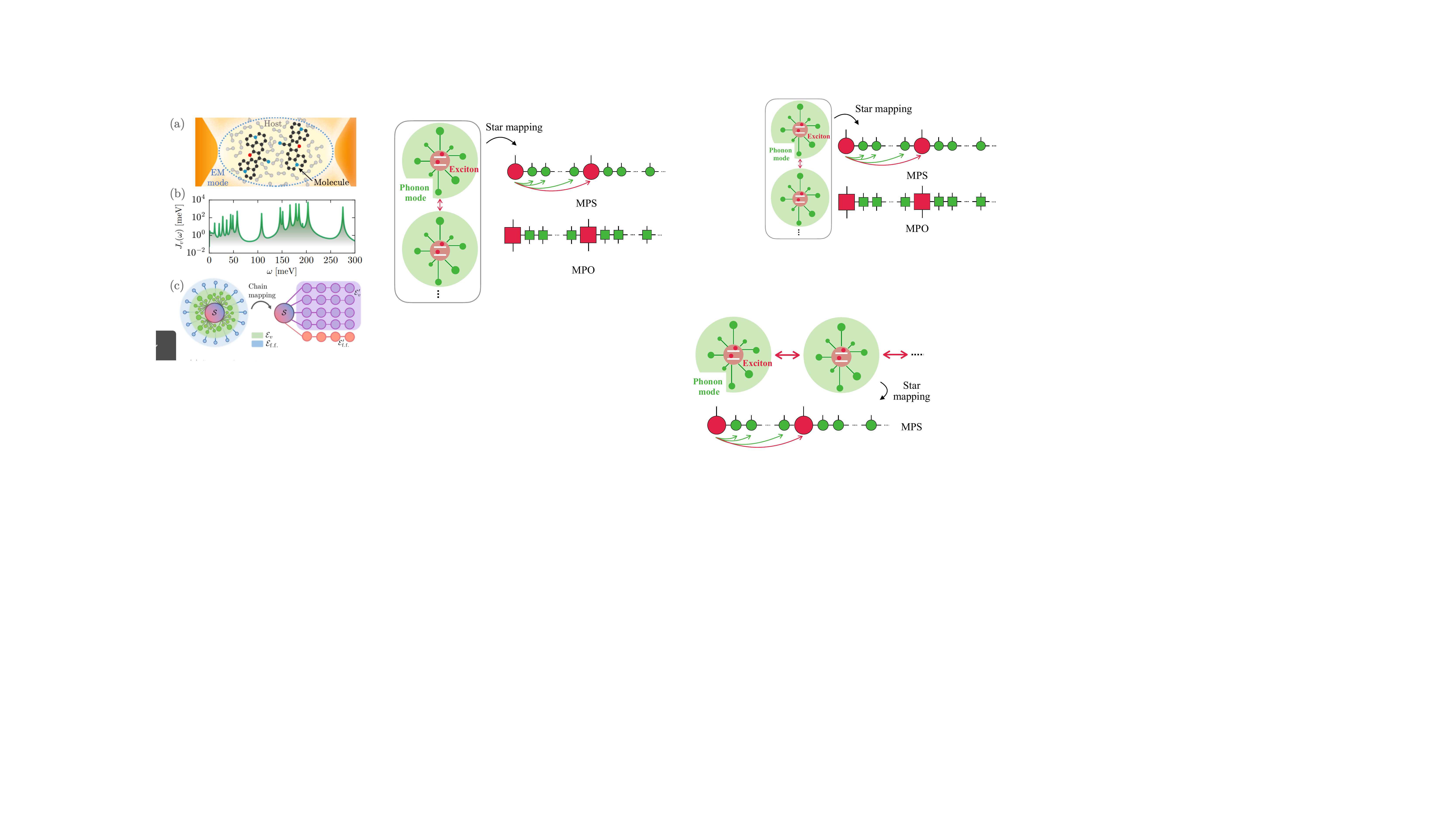}
\caption{Scheme illustrating the mapping of the vibrational modes yielding the MPS representation of states/operators used in the simulations.}
\label{fig:scheme}
\end{figure}
Our theoretical analysis for aggregate excitonic structure and spectroscopy is based on the following Frenkel-Holstein model:
\begin{eqnarray}
        \hat{H} = &&\sum_{m,n}^N J_{mn} a_m^\dagger a_n  
        + \sum_{n}^{N}\sum_{i}^{N_{\textrm{ph}}} \omega_{ni} b_{ni}^\dagger b_{ni}\nonumber\\
        &&+ \sum_{n}^N\sum_i^{N_{\textrm{ph}}} \omega_{ni} g_{ni} a_n^\dagger a_n (b_{ni}^\dagger + b_{ni}),
        \label{eq:model}
\end{eqnarray}
in which $N$ molecules are coupled to a set of $N_{\textrm{ph}}$ phonon modes with frequency $\omega_{ni}$ and coupling strength $g_{ni}$. $S_{nj}=g_{nj}^2$ is also known as the Huang-Rhys factor.
The excitonic coupling $J_{mn}$ describes the transition dipole-dipole interaction between molecules ($m\neq n$).
The on-site transition energy is set to be 0.
$a_n^\dagger$ ($a_n$) is the raising (lowering) operator for local excitation, and $b^\dagger_{ni}$ ($b_{ni}$) is the phononic creation (annihilation) operator associated to the $i$th phonon mode coupled to the $n$th molecule.
The thermal equilibrium density matrix $\rho_{\beta}$ of a mixed state in physical space $ P $ can be obtained by performing the partial trace of the pure state in the double space $P\otimes Q$,
\begin{equation}
\hat{\rho}_{\beta}=\frac{e^{-\beta \hat{H}}}{Z}=\frac{\textrm{Tr}_Q|\psi_\beta\rangle\langle\psi_{\beta}|}{\textrm{Tr}_{PQ}|\psi_\beta\rangle\langle\psi_{\beta}|},|\psi_{\beta}\rangle=e^{-\beta\hat{H}/2}|I\rangle.
\end{equation}
where $\beta=1/k_\textrm{B} T$. The many-body wavefunction $|\psi_{\beta}\rangle$ in the enlarged space is approximated as MPS, as shown in Fig.~\ref{fig:scheme}. It is obtained by the propagation of the maximally entangled state $|I\rangle=\sum_i|i\rangle_P |i\rangle_Q$ along the imaginary time axis to $-\beta/2$ employing the time dependent variational principle with projector splitting algorithm~\cite{haegeman2016unifying}.
The radiative efficiency of $N$ site molecular aggregate is defined as the ratio between its oscillator strength over that of an isolated molecule, 
\begin{equation}
\gamma=\frac{\int d\omega S_N(\omega)}{\int d\omega S_1(\omega)}
\end{equation}
\begin{equation}
S(\omega)=\sum_{m,n}\frac{e^{-\beta E_n}}{Z}|\langle m|\hat{\mu}|n\rangle|^2\delta(\omega-E_{mn})\label{eq:freq}
\end{equation}
where $S(\omega)$ is the fluorescence spectroscopy. The transition dipole operator $\hat{\mu}=\mu\sum_{i}^N(a_i^\dagger +a_i)$, $E_{mn}=E_m-E_n$.
The integration of $S(\omega)$ over all frequency is equal to the oscillator strength (note that we neglect the minor cubic dependency about frequency of $S(\omega)$).
In cases of the nonradiative decay rate much slower than radiative rate, $\gamma$ approximates the quantum yield.
The radiative efficiency is directly related with ECL which depicts the number of coherently coupled excitons.
In the complete coherence limit, the ECL is equal to the total number of molecules $N$, and the instructive oscillations of dipoles leads to $N$ times superradiance enhancement in J agggregate or the destructive oscillations of dipoles leads to complete quench in H aggrgegate, while in the complete localization limit, the ECL equals 1, namely,
the radiative efficiency of the molecular aggregate is equal to that of the monomer.
Therefore, a certified ECL bridges the radiative efficiency and coherence of molecular aggregate.

Different ECLs have been defined by:
\begin{align}
	&L_1=\frac{(\sum_{mn}|\rho_{mn}|)^2}{N\sum_{mn}|\rho_{mn}|^2}, \quad 1\le L_1\le N\label{eq:variance}\\
	&L_2=\sum_{m\ge n}|\rho_{mn}|, \quad 1\le L_2 \le\frac{N+1}{2}\label{eq:adds}\\
	&L_3=N\textrm{Tr}\rho^2, \quad 0\le L_3\le N\label{eq:purity}
\end{align}
where $L_1$ corresponds to the variance of the reduced density matrix, $L_2$ sums the absolute values of the upper triangles of the reduced density matrix, and $L_3$ makes use of the purity of density matrix ($\textrm{Tr}\rho^2\le 1$, the equivalence holds for pure state).
All these definitions give monotonic relationship between the exciton coherence length and radiative efficiency in the case of purely excitonic systems (Fig.S1), $\textit{ie.}$, without considering the interplay of phonons. 
After the exciton-phonon couplings are involved, we calculate the reduced density matrix $\rho$  by tracing the phonon degrees of freedom of the total Boltzmann operator
\begin{equation}
\rho=\textrm{Tr}_{\textrm{ph}}\hat{\rho}_{\beta}
\end{equation}
with elements $\rho_{mn}=\textrm{Tr}(a_m^\dagger a_n\hat{\rho}_{\beta})$ containing the coherence between molecule $m$ and $n$.
\begin{figure}
\centering
\includegraphics[width=0.5\textwidth]{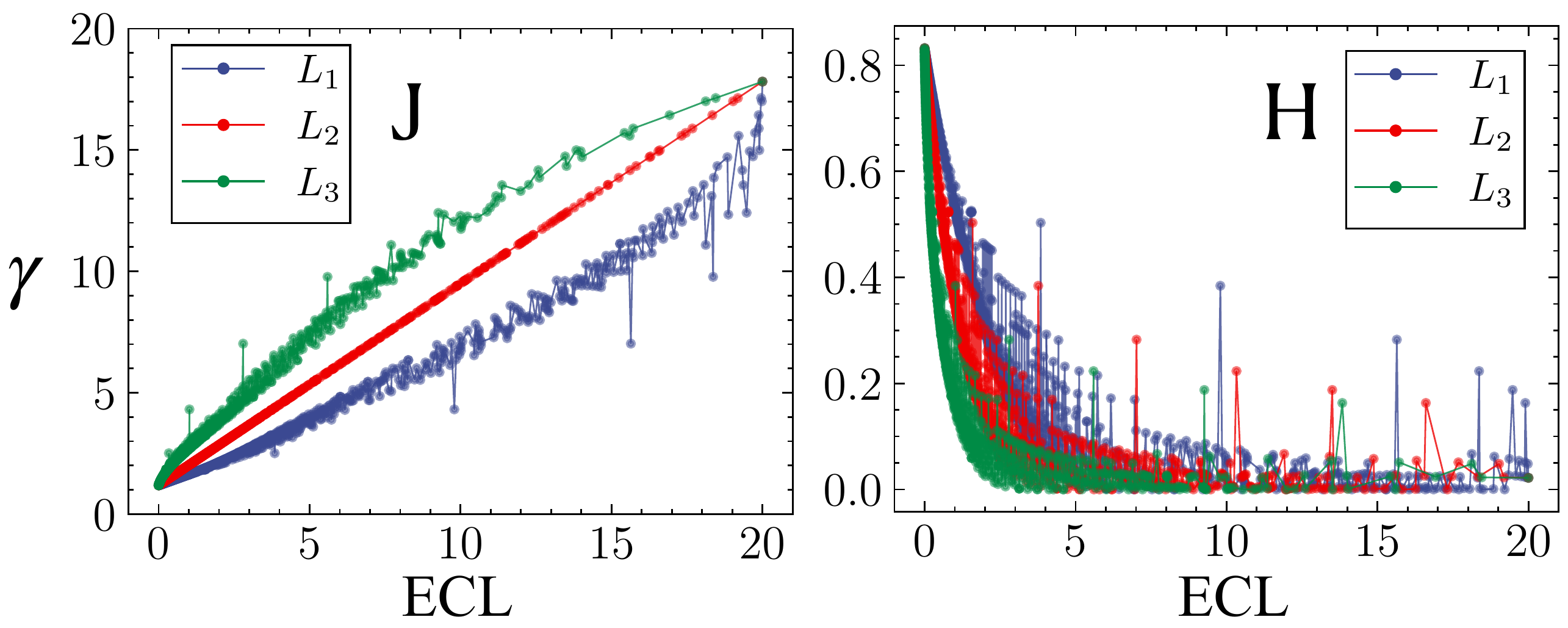}
\caption{The relationship between the radiative efficiency and ECL with different definitions for 1d J aggregate and H aggregate.}
\label{fig:ECL_1d}
\end{figure}

Firstly, we consider homogeneous 1d J and H aggregate made of interacting dipoles with periodic boundary condition (PBC), as shown in Fig.~\ref{fig:ECL_1d}.
The benchmarks are shown in Fig.S2.
Each molecule couples to two phonon mode: a high frequency mode with frequency $\omega$ and Huang-Rhys factor $S_1=0.5$, that represents the ubiquitous vinyl stretching/ring breathing vibration observed in most conjugated molecules, and a low frequency mode with frequency $0.1\omega$ that corresponds to such as torsional motions.
We plot the relationship between ECLs with different definitions and $\gamma$ for J aggregate and H aggregate in  Fig.~\ref{fig:ECL_1d}.
Among the parameters studied in Fig.~\ref{fig:ECL_1d}, the excitonic coupling is adjusted from $0.1\omega$ to $\omega$, and the Huang-Rhys factor of the low frequency mode is adjusted from $0$ to $5.0$, and the temperature is adjusted from 0 to $2.0\omega$. The chosen parameters ensure a broad parameter regime that covers from localization limit to delocalization limit.
The absolute values of ECLs are all scaled to the window of $L_1$ since the maximal and minimal value for different definitions are different (see Eq.~\eqref{eq:variance}~\eqref{eq:adds}\eqref{eq:purity}), and we particularly interested in the ability of ECLs of reflecting the changes of radiative efficiency.
As shown in Fig.~\ref{fig:ECL_1d}, nonmonotonous and multiple-valued correlations are observed for both J and H aggregate, in contrast to the purely excitonic case presenting monotonic relations (Fig.S1). 
We also note the inappropriate faster growing ECL of $L_1$ near the limit of complete coherence, which results in the case that systems have similar ECLs but vastly different radiative efficiency (Fig.S3).

\begin{figure}
\centering
\includegraphics[width=0.45\textwidth]{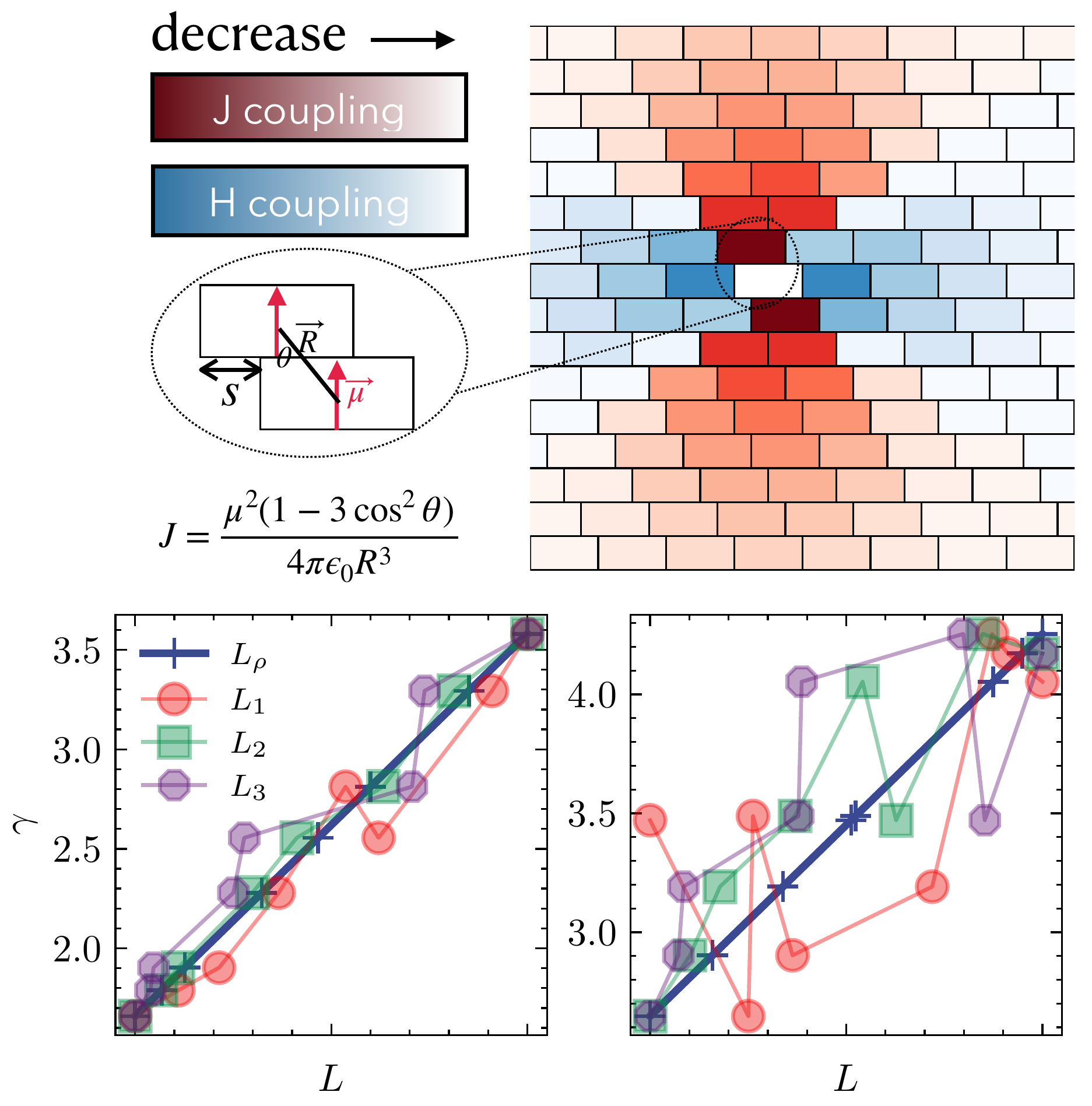}
\caption{(a) Illustration of the 2d brick-layer system with red/blue colors represent positive/negative excitonic couplings with the centered molecule (in white), the length $l$ and width $l/2$ are set for each brick. The correlation between radiative efficiency $\gamma$ and ECL with different definitions after changing the slip distance $0\le s< l$ at temperature $T=0.5\omega$ for (b) $J_0=\omega$ and (c) $J_0=2\omega$ for $10\times 10$ bricks with open boundary condition. $J_0$ is the nearest-neighbor coupling along the vertical axis for $s=0$. The phonon mode is with frequency $\omega$ and $g=1.$ $M=64$ is used.}
\label{fig:brick}
\end{figure}

Because of the failure of the available ECL definitions for 1d H aggregate, they are generally not suitable to be used in the common real-world systems with  2d stacking structures, one defining feature of  which is the coexistence of both positive and negative excitonic couplings (Fig.~\ref{fig:brick}).
We model 2d aggregates packed with brick-wall structures~\cite{chuang2019generalized} with inhomogeneous excitonic couplings (\textit{i.e.}, the couplings change strengths or even signs for different pairs of dimers).
The exciton systems are with planer brick layer structures, and the change of geometry by adjusting the slip distance between neighboring horizontal layers leads to different excitonic coupling patterns, as shown in Fig.~\ref{fig:brick}.
The couplings are distinguished by red for positive coupling and blue for negative coupling, and the color depth indicates the strength of excitonic couplings.
The excitonic coupling between two bricks $J=\mu^2(1-3\cos{\theta}^2)/4\pi\epsilon_0 R^3$, where $\vec{R}$ is the displacement vector between the brick centers and $\theta$ is the angle between $\vec{R}$ and the transition dipole vector $\vec{\mu}$, and $\epsilon_0$ is the dielectric constant in the medium.
In Fig.~\ref{fig:brick}, we change the coupling strength by adjusting the nearest-neighbor coupling $J_0$ for s=0, $\textit{ie.}$, $J_0=\frac{\mu^2}{2\pi\epsilon_0 w^3}$ where $w$ is the brick width.
By varying the slip distance $s$ ($0\le s \le 0.875l$), we observe the change of radiative efficiency and coherence length with different definitions, as shown in Fig.~\ref{fig:brick}(b)(c).
All values of ECL with different definitions are scaled to the same window for comparison.
As shown in Fig.~\ref{fig:brick}, in the weak excitonic coupling case, the exciton coherence lengths are all well overlapped with the radiative efficiency as the slip distance increases.
However, for larger excitonic couplings, previously defined exciton coherence lengths display even qualitative differences.

\begin{figure}
\centering
\includegraphics[width=0.48\textwidth]{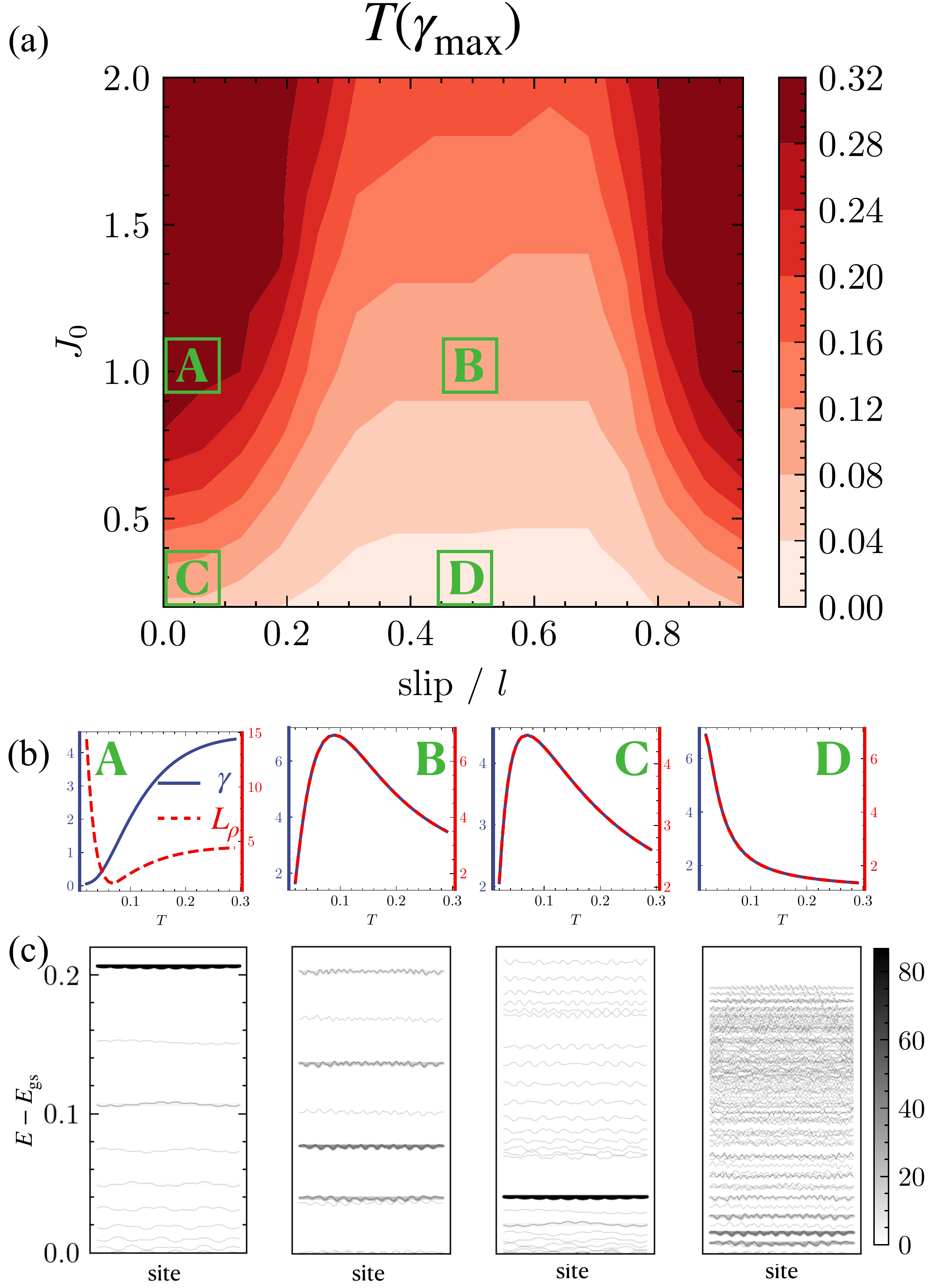}
\caption{(a) The temperature for maximum superradiant enhancement for different slip distances and excitonic couplings. (b) Four exemplary cases that illustrates the temperature dependence of $\gamma$ and $L_\rho$ and (c) the corresponding wavefunctions with combining coefficients for local site, whose color depth indicating the oscillator strength of that energy level.}
\label{fig:prediction}
\end{figure}

Against the fact that previously proposed ECLs cannot correlate well with radiative efficiency, we propose the following definition for ECL:
\begin{equation}
L_{\rho}=(\sum_{mn}^N\rho_{mn}+\frac{1}{N})^{\Theta(\sum_{mn}^N\rho_{mn}-1)}\label{eq:coh}
\end{equation}
where $\Theta(x)$ is a step function that is -1 for $x\le 0$ and 1 for $x>0$.
This function builds single-valued correlation between ECL and radiative efficiency (see Fig.S4) since it utilizes the fact that the summation of all elements of the reduced density matrix is equal to the radiative efficiency $\gamma$.
It is straightforward to check the equivalence between $\gamma$ and $\sum_{mn}^N\rho_{mn}$, which is a kind of sum rule $\int d\omega S_N(\omega)=\textrm{Tr}(\hat{\rho}_{\beta}\hat{\mu}\hat{\mu})$.
The computational time can also be greatly saved since it is unnecessary to perform dynamical response calculation (Eq.~\eqref{eq:freq}) to calculate $\gamma$.
It is useful to analyze the limiting cases. In the complete coherence limit, 
\begin{equation}
\rho_{mn}=1/N, L_{\rho}=N+1/N\sim N\quad \textbf{J aggregate}
\end{equation}
\begin{equation}
\rho_{mn}=(-1)^{m-n}/N, L_{\rho}=N\quad \textbf{H aggregate}
\end{equation}
In the complete localization limit for both J and H aggregate, 
\begin{equation}
\rho_{mn}=\delta_{mn}/N,L_{\rho}=N/(N+1)\sim 1.
\end{equation}
Starting from the reduced density matrix, a determined value of exciton coherence length is connected with a given radiative efficiency, which is different from previous definitions.
Such definition ensures linear relationship between coherence and radiative efficiency for the J aggregate case, which is also related with the concept of quantum Fisher information that measures the multipartite entanglement in quantum metrology of exciton system~\cite{sifain2021toward}.
As shown in Fig.~\ref{fig:brick}, $L_\rho$ correlates linearly with the radiative efficiency $\gamma$ (Fig.S4).

We further study the temperature dependence of the superradiance enhancement (SRE) of the 2d brick layers.
The temperature dependence of SRE was described by $1/T$ law~\cite{spano2004temperature,arias2013thermally}, namely,
the temperature achieving maximal SRE is zero, and the SRE decreases with increasing temperature.
Here, we find absolute new temperature dependence of the SRE, as shown in Fig.~\ref{fig:prediction}.
We observe nonmonotonous temperature dependence of radiative efficiency in the 2d agggregates made of brick layers and realizing maximal SRE at finite temperature ($T(\gamma_{\textrm{max}})>0$) is possible, in contrary to the previous $1/T$ behavior.
We plot the temperature when achieving maximal SRE when choosing different excitonic coupling strengths and slips in Fig.~\ref{fig:prediction}(a).
The temperature dependence experiences a transition when temperature increases to a certain point $T(\gamma_{\textrm{max}})$, after which the SRE decreases with increasing temperature.
It is shown that $T(\gamma_{\textrm{max}})$ becomes larger for larger excitonic coupling.
We select 4 exemplary cases (A, B, C, D) in Fig.~\ref{fig:prediction}(a)(b) with different temperature dependences for $\gamma$ and $L_\rho$. 
The corresponding excitonic coupling patterns are shown in Fig.S5.
For case A, the SRE keeps increasing with the increase of temperature, which is totally different from traditional J aggregate whose SRE decreases as temperature increases.
In traditional J aggregate, most of the oscillator strengths are located in the lowest excited state, and the increased temperature reduces the thermal population on the lowest state, leading to decreased SRE.
While in traditional H aggregate, the oscillator strengths are located in the highest excited states, thus the increased temperature increases the population on the highest states, leading to increased radiative efficiency but will not exceed 1.
Case A totally violates the temperature dependence of the radiative efficiency of traditional J and H aggregate: it is superradiant as J aggregate but the radiative efficiency grows with increasing temperature, just like H aggregate.
Case D behaves as traditional J aggregate.
By contrast, the SRE of case B and C first increases with temperature then decreases.
The differences are resulted from the differences of corresponding energy structures, we explain this by directly diagonalizing the exciton Hamiltonian and get the wavefunction amplitudes of each site and the oscillator strength of each energy level in Fig.~\ref{fig:prediction}(c).
The energy of ground state energy is set as the zero point.
For case A, the energy levels with energy below 0.2 are almost dark states, while the oscillator strength concentrates on the high energy level, therefore,
the radiative efficiency keeps increase with increasing temperature since the occupation of the bright state becomes larger.
At very low temperature of case A, the exciton is distributed mainly at the low energy levels which is very delocalized but with destructive wavefunction amplitudes. 
It is observed that the radiative efficiency is lower than 1, which is the characteristic of H aggregate. As temperature increases, the occupation of the lowest levels becomes smaller, and the coherence decreases until the radiative efficiency grows to 1 which marks the transition to J aggregate.
The occupation of the high coherent level with large oscillator strengths becomes larger, and the coherence length and radiative efficiency increases with temperature.
In contrary to case A, the SRE and coherence length of case D decreases with increasing temperature since the coherent bright state is located at the bottom.
The coherent bright states of case B and C are located neither in the middle of energy diagrams, therefore, the SRE experiences a thermal activation at low temperature, after which it decreases with increasing temperature. 

In summary, we provided a unified definition of exciton coherence length (ECL) for molecular aggregates. Previously, three definitions that employ the variance, off-diagonal elements, or the purity of reduced density matrix cannot correlate ECL with the radiative efficiency (RE) in a single-valued way, and even leading to difference trends. Our new definition is derived by virtue of sum rule of oscillator strength. Using the numerically accurate time-dependent matrix product states applied to both one-dimensional and two-dimensional Frenkel-Holstein models for both H- and J-type molecular aggregates, we find our ECL definition rigoroursly connect with the radiative efficiency of exciton-phonon systems, and serves as an efficient and reliable quantity for exciton coherence measurement.
Our results suggested new temperature dependence of superradiance and coherence in 2d aggregates with brick-wall packing structures,
which can be used for the design of light emitting materials, since they demonstrate that the radiative efficiency can be dramatically adjusted by the relative slips between layers of 2d molecular aggregates.

This work is supported by the National Natural Science Foundation of China (NSFC) through the project ``Science CEnter for Luminescence from Molecular Aggregates (SCELMA)'' Grant Number 21788102 and through Grant Number 22003029, as well as by the Ministry of Science and Technology of China through the National Key R\&D Plan Grant Number 2017YFA0204501.
\bibliographystyle{apsrev4-2}
\bibliography{ref}
\end{document}


\title{Supplemental materials}
\maketitle

\renewcommand{\theequation}{S\arabic{equation}}
\renewcommand{\thefigure}{S\arabic{figure}}
\renewcommand{\bibnumfmt}[1]{[S#1]}
\renewcommand{\citenumfont}[1]{S#1}
\begin{figure}
\centering
\includegraphics[width=0.8\textwidth]{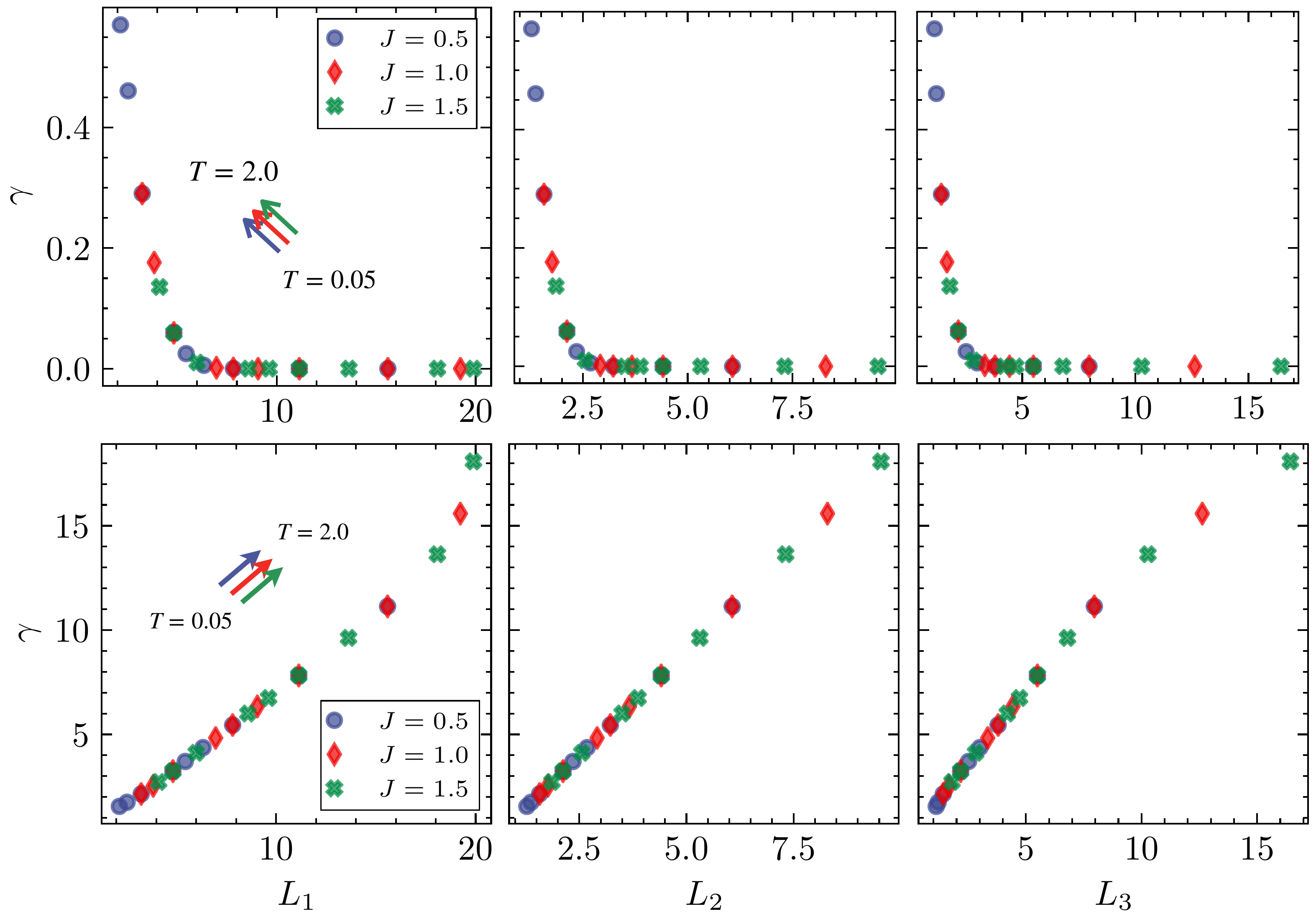}
\caption{The relationship between the radiative efficiency $\gamma$ and exciton coherence length $L_1, L_2, L_3$ for 20-site disorder-free pure excitonic system. The periodic boundary condition (PBC) is employed.}
\label{fig:S1}
\end{figure}

\begin{figure}
\centering
\includegraphics[width=0.8\textwidth]{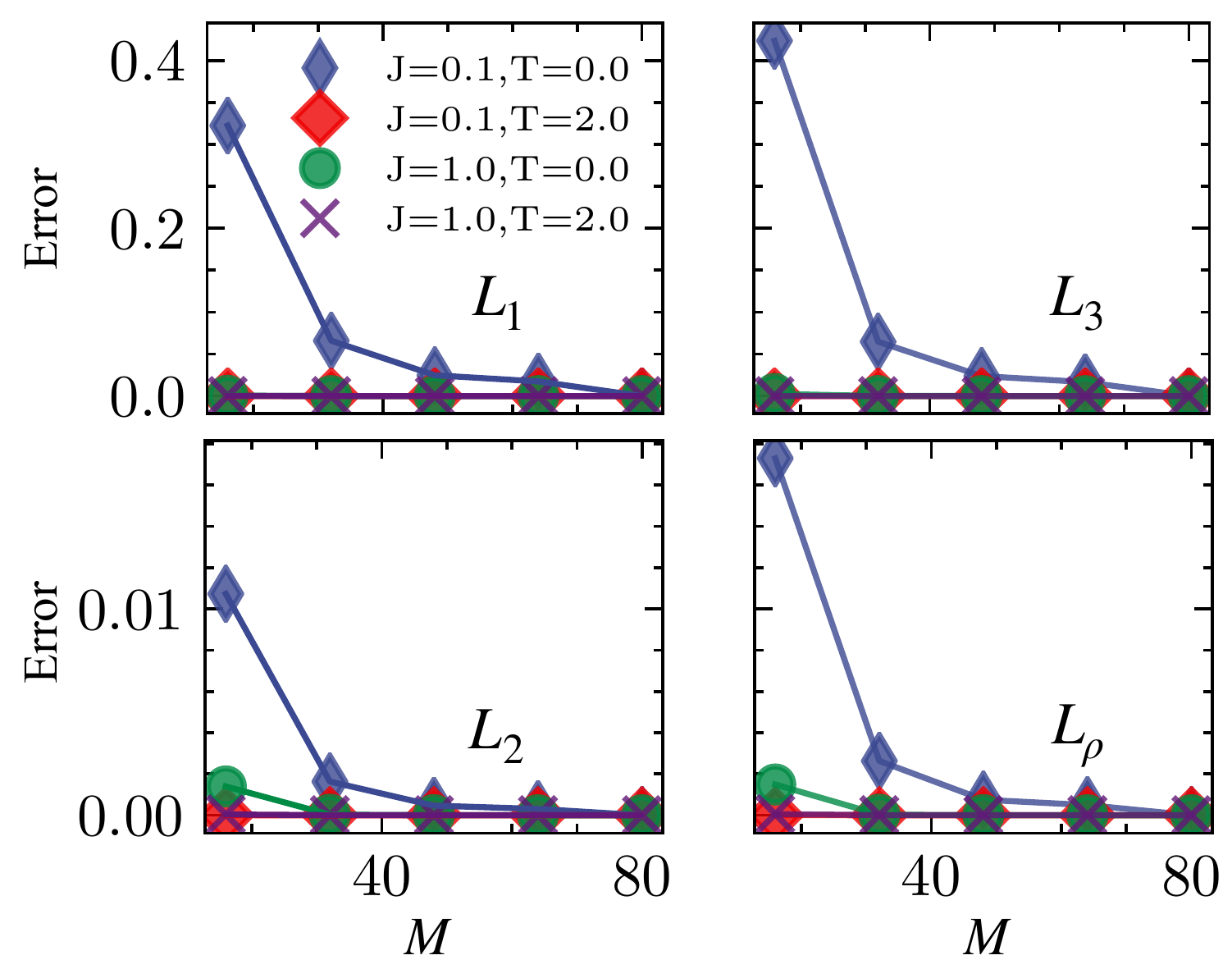}
\caption{Convergence of ECLs versus the bond dimension $M$ for limiting cases of 20-site 1d aggregate with PBC. The relative error is computed with reference of $M=80$. The phonon mode is chosen with frequency $\omega=1$, $S_1=0.5$, and $\omega_2=0.1$, $S_2=5.0$.}
\label{fig:bench1d}
\end{figure}

\paragraph{Benchmarks.}We benchmark the ECLs at limiting cases of one-dimensional (1d) aggregates, as shown in Fig.~\ref{fig:bench1d}. 
The phonon modes are chosen with high frequency $\omega=1$ with Huang-Rhys factor $S_1=0.5$, and low frequency $\omega_2=0.1$ with $S_2=5.0$.
The temperature and excitonic couplings are adjusted.
We find that the convergence of bond dimension $M$ of the matrix product state is slow for the low temperature limiting case with strong exciton-phonon couplings for all definitions, otherwise the convergence is quick.
We also find that the convergences of $L_1$ and $L_3$ at strong exciton-phonon coupling cases at low temperature is harder.
If we let $S_2=10.0$, the convergence of $L_1$ and $L_3$ is very difficult for $J=0.1$ at zero temperature, which is not reached even increasing $M=128$. By contrast, $L_2$ converge much easier with $M=32$.

\paragraph{Problems of $L_1$.} The exciton coherence length calculated with definition 1 is $L_{1,|J|=0.1}=18.7$ and $L_{1,|J|=1.0}=19.8$ for the case of weak excitonic coupling and the case of strong coupling respectively, which are close to each other and imply nearly complete coherence for both cases. 
However, the radiative efficiency $\gamma$ in strong coupling case is $1.6$ times larger than the weak coupling case for J aggregate. The difference is more pronounced for H aggregate that exhibit $0.1$ times suppression of $\gamma$. Namely, the values of radiative efficiency and coherence length $L_1$ is inconsistent.
\begin{figure}
\centering
\includegraphics[width=0.8\textwidth]{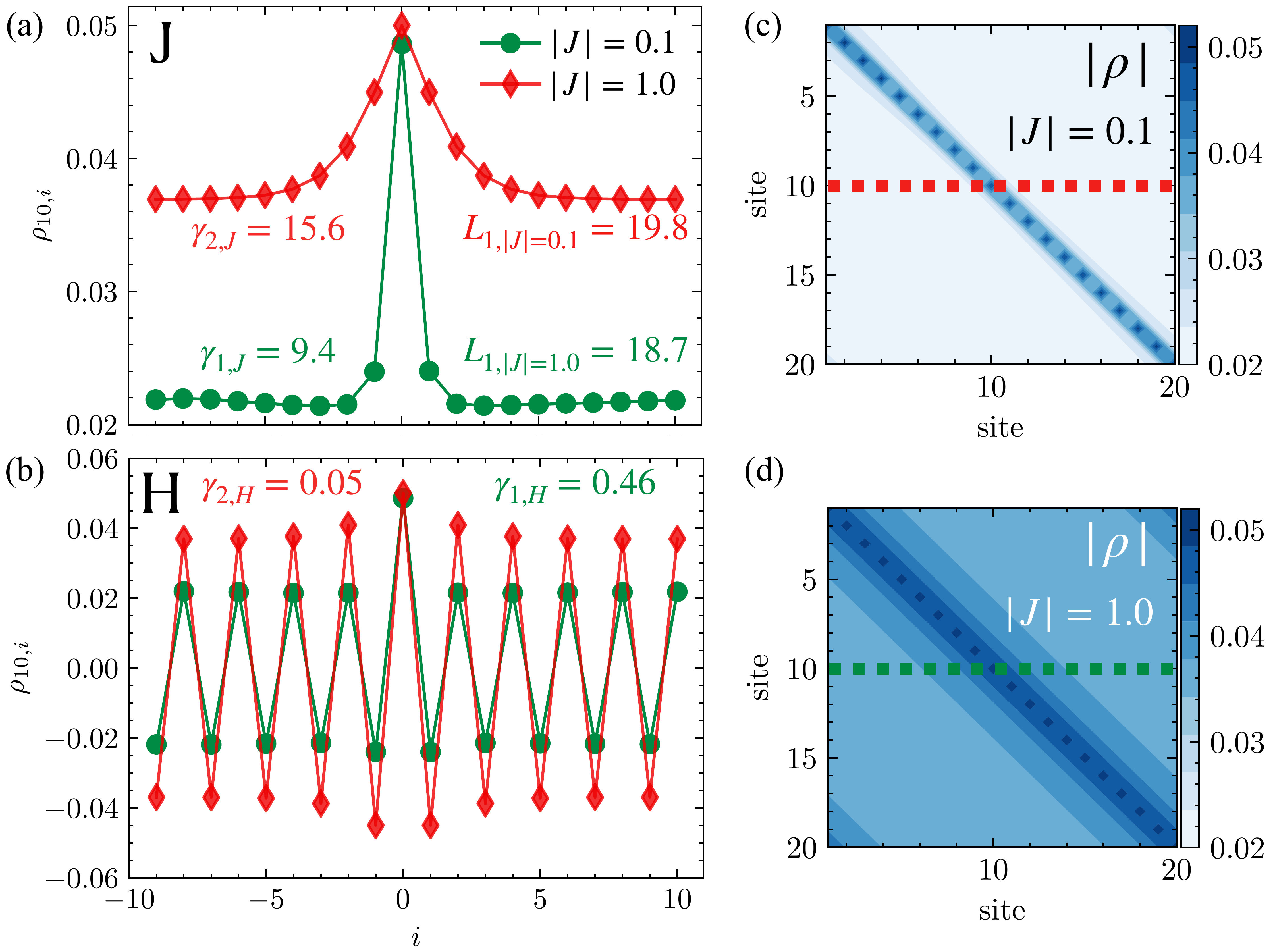}
\caption{The reduced density matrix information of $N=20$ linear chain with periodic boundary condition (PBC). The reduced density matrix elements for the $10$th molecule, $\rho_{10, i}$ for (a) J aggregate ($J<0$), and (b) H aggregate ($J>0$) with two different absolute values of $J$. The absolute values of entire $\rho$ for (c) $|J|=0.1$ and (d) $|J|=1.0$ in unit of $\omega$. }
\label{fig:outlier}
\end{figure}
\begin{figure}
\centering
\includegraphics[width=0.8\textwidth]{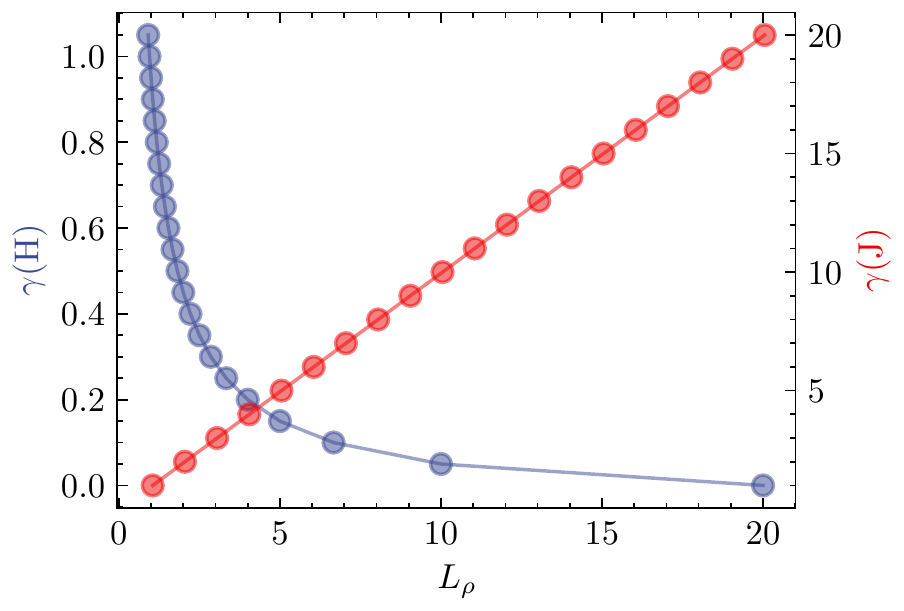}
\caption{The relationship between the radiative efficiency $\gamma$ and exciton coherence length $L_\rho$ for exciton-phonon system at finite temperature. The periodic boundary condition (PBC) is employed.}
\label{fig:mono}
\end{figure}
\begin{figure}
\centering
\includegraphics[width=0.8\textwidth]{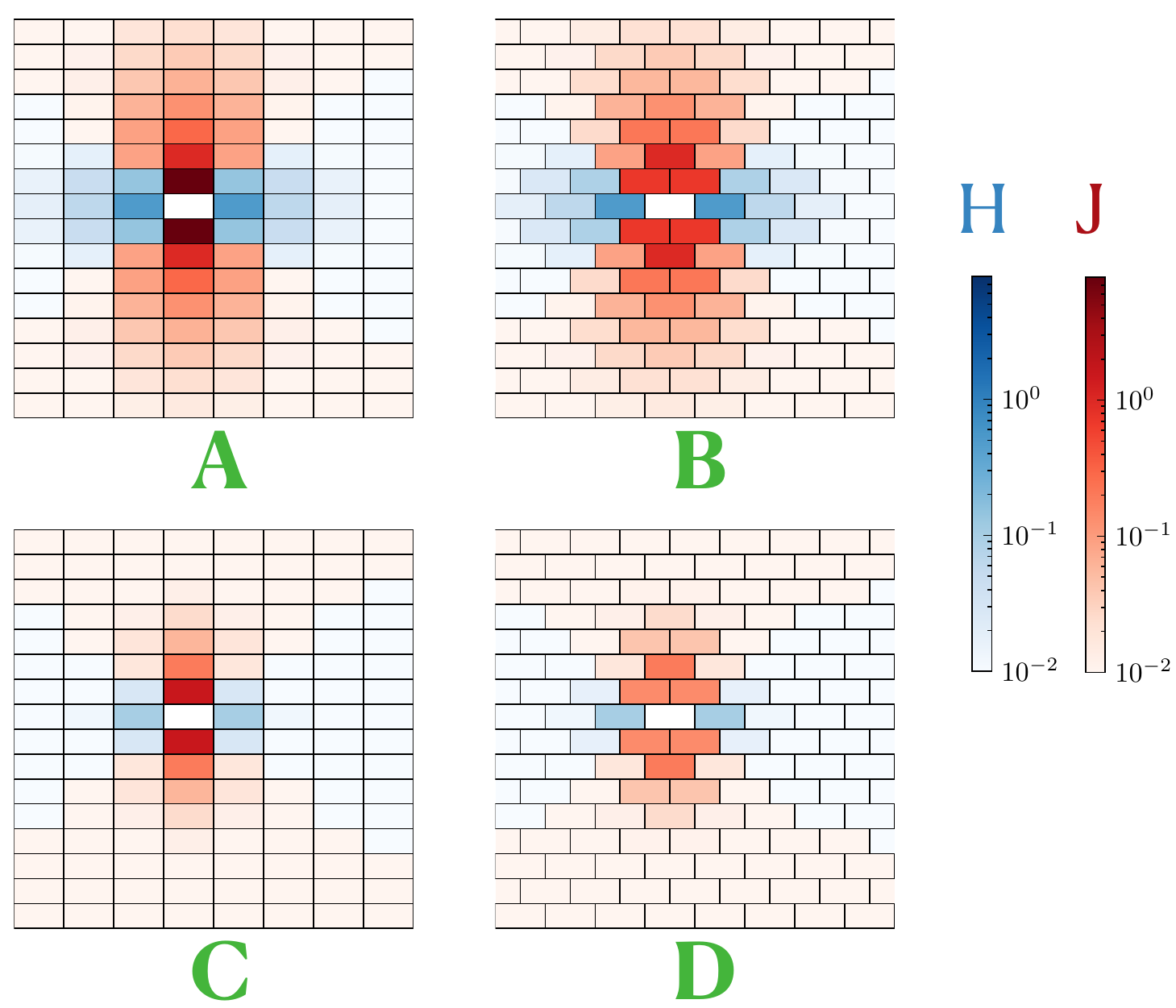}
\caption{The corresponding excitonic coupling patterns for the exemplary cases in Fig.4.}
\label{fig:coup_pattern}
\end{figure}